\documentclass[preprint,aps,showpacs,showkeys]{revtex4}%
\usepackage{amsfonts}
\usepackage{amsmath}
\usepackage{amssymb}
\usepackage{graphicx}%
\providecommand{\U}[1]{\protect\rule{.1in}{.1in}}

\begin{document}
\title[ ]{Submillimeter constraints for non-Newtonian gravity from spectroscopy }
\author{A. S.\ Lemos}
\email{adiellemos@gmail.com}
\affiliation{Departamento de F\'{\i}sica, Universidade Federal de Campina Grande - Campina
Grande - PB - Brazil}

\begin{abstract}
In this work, we consider the Yukawa-type and power-type non-Newtonian
corrections, which induce amplification of gravitational interaction on
submillimeter scales, and analytically calculate deviations produced by the
atomic gravitational field on the energy levels of hydrogen-like ions.
Analyzing ionic transitions between Rydberg states, we derive prospective
constraints for non-Newtonian corrections. It is shown that the results also
provide stronger constraints, due to the high accuracy for Rydberg
transition measures into optical spectrum frequency range, than the current
empirical bounds following from Casimir force measurements.

\end{abstract}
\maketitle

\section{Introduction}

Several theoretical scenarios in different contexts have predicted possible
deviations from Newtonian gravity at short-distance scales \cite{fischbach1}.
Their initial formulation was invoked for the fifth force theory, known for
predicting a new long-range force that would influence the interactions
between the particles up to a certain distance \cite{fischbach2}. In this
framework, the detection of corrections to Newton's inverse square law of
gravity, for instance, would imply the existence of a fifth force in addition
to known interactions predicted by the Standard Model
\cite{fischbach1,fischbach2,adelberger}.

In its turn, direct tests of the non-Newtonian gravity have attracted
attention in recent years due to the predictive accuracy of these theories,
compared to the current level of experimental uncertainty, and also motivated
by the possibility of find traces of new physics
\cite{hoyle2001,hoyle2004,hoyle2007}. Recent experiments using torsion-balance
in a regime of submillimeter scales have attempted to examine since the
validity of some physical principles, such as the Weak Equivalence Principle,
e.g., see \cite{wep}, even the existence of a dark-energy length scale as the
regulator of a new fundamental length scale for gravity \cite{hoyle2007}. In
this case, the results impose limits on corrections to Newton's gravity law,
showing the absence of deviations up to a submillimeter scale
\cite{murata,Antoniadis}.

From measuring the Casimir force, constraints to predicted deviations to
Newtonian gravity have also been extracted
\cite{murata,mostepanenko,mostepanenko2}. In this context, the non-Newtonian
interaction is assumed to present Yukawa-type, and power-type corrections
arising from the exchange of particles, such as light scalar particles
\cite{mostepanenko,mostepanenko2}.

In another way, there is a possibility that the non-Newtonian theory of
gravity can be described as a theory of gravitation modified by the existence
of extra dimensions \cite{murata}. Therefore, the Newtonian gravitational
potential must present corrections on short-range scales due to compact extra
dimensions, whose compactification radius $R$ is of the sub-micrometer order.
However, the four-dimensional behavior of gravitational potential is recovered
if the distances regime $r$ is greater than the radius $R$ \cite{add1,add2}.
In this theoretical framework, the usual Newtonian gravitational potential has
a Yukawa-type correction, for instance, that presents an explicit dependence
with the space-like dimensionality and the compactification scale
\cite{kehagias}. Hence, if experimental deviations are measures, the fifth
force could be evidence for empirical traces from\ extra dimensions.

In the scenario of extra-dimensional theories proposed by unification schemes,
one can highlight the known braneworld models, which have an elegant geometric
description of the ordinary universe inspired by string theories. According to
this theoretical model, our $4$-dimensional observable universe is a
submanifold, in which the standard model fields are confined, embedding in a
higher-dimensional space with $\delta$ large extra dimensions $\left(  R\gg
l_{Planck}\right)  $ \cite{add1,add2}. In this context, the spectroscopy has
proved as an independent way to test deviations from the inverse-square
law
\cite{lemos1,lemos2,lemos3,lemos4,specforxdim1,specforxdim2,molecule,safranova,ryd}%
. For instance, in ref. \cite{lemos1}, there have been obtained stronger
constraints to compactification radius and higher-dimensional Planck mass from
analyzing the $1s-2s$ electronic transition in the hydrogen by considering the
thick braneworld scenario. Furthermore, in recent work \cite{lemos2}, it was
shown that extra dimensions with a thickness ($\sigma\lesssim10^{-19}m$) could
provide the excess energy found in the recent measurements of muonic hydrogen
$2S-2P$ transitions, and so solve the known proton radius puzzle
\cite{nature,science}. Thus, the proton radius puzzle would signal new physics
coming from large extra dimensions, which would imply a violation of Newton's
gravity inverse-square law \cite{lemos2}. Moreover, recently the spectroscopy
data analysis of the hydrogen-like atoms in Rydberg states has been applied to
study the influence of a new Yukawa-type interaction on energy levels, and the
atomic spectra deviations have been obtained \cite{ryd}.

This work aims to discuss prospective constraints on deviations from the
non-Newtonian gravitational law of power-type and Yukawa-type from
spectroscopic data and compare them. This paper is structured as follows: In
the next section, we present the power-type correction to Newton's gravity law and
explicitly calculate the gravitational contribution for the ionic energy
levels. Then, considering states with high angular momentum, we study specific
transitions in the neon electronic and muonic atoms aiming to obtain
constraints on the parameters of this model by assuming that the empirical
predicted uncertainty to transitions localizing in the optical frequency
spectrum limits the found deviations. In the third section, we consider the
non-Newtonian potential with a Yukawa-type correction and calculate
gravitational energy contribution to transitions between Rydberg states for
the electronic and muonic hydrogen-like ions. From this analysis, we have
presented constraints to the theoretical model and discuss the results. The
bounds have been obtained requiring that the gravitational energy contribution
does not exceed the promised empirical uncertainty, considering the leading
term of atomic Hamiltonian. In its turn, in the fourth section, we have compared the
deviations on the atomic energy spectra generated by both proposed
non-Newtonian parameterizations. Finally, in the last section, we present the conclusions.

\section{Power-law non-Newtonian potential parametrization}

Inspired by the proposed Arkani-Dimopoulos-Dvali (ADD) model of large extra
dimensions, one can write the corrections to Newtonian gravitational potential
due to higher dimensionality of spacetime according to the following equation
\cite{add1,add2},%

\begin{equation}
V\left(  r\right)  =\left\{
\begin{array}
[c]{c}%
-\frac{G_{4+\delta}M}{r^{1+\delta}},\text{\hspace{0.8cm}if }r\ll\lambda\\
-\frac{G_{N}M}{r},\text{\hspace{1cm}if }r\gg\lambda
\end{array}
\right.  , \label{add}%
\end{equation}
where $\lambda$ is the compactification scale related to the extra dimension
radius. In this case, we have assumed that the spacetime has $\delta$
extra dimensions.

The ($4+\delta$)-dimensional gravitational constant, $G_{4+\delta}$, can be
easily rewritten in terms of the usual $4$-dimensional Newtonian constant,
$G_{N}$, by assuming that at regime $r\gg\lambda$, we should recover the
four-dimensional behavior for gravitational potential. Thus, we find%

\begin{equation}
G_{4+\delta}=\lambda^{\delta}G_{N}.
\end{equation}

Conveniently, let us express the eq. (\ref{add}) through a function that
interpolates and describes the Newton's gravitational potential behaviors into
two regimes ($r\ll\lambda$ and $r\gg\lambda$). Thus, the power-type
parametrization which generalize the gravitational potential can be obtained
directly \cite{murata}:%

\begin{equation}
V\left(  r\right)  =-\frac{G_{N}M}{r}\left(  1+\left(  \frac{\lambda}%
{r}\right)  ^{\delta}\right)  , \label{power potential}%
\end{equation}
where $\delta=0,1,2,\ldots,$ $\lambda$ is an interaction constant associated
with the extra-dimensional radius and, consequently, has length dimension.
From experimental data will constrain both parameters. Let us mention
that several extensions of General Relativity also predict this
parametrization and even $f\left(  R\right)  $ theories \cite{peri}.

\subsection{Shifted non-Newtonian gravitational atomic energy}

At this point, we are interested in analyzing transitions between Rydberg
states with high angular momentum. Therefore, the effects of the nuclear
structure are negligible, and the uncertainty originated from the proton radius is
small, so we do not take into account the fact that the proton has a finite
structure \cite{lemos3}. Hence, in the first approach, the atomic energy
levels will be perturbed by gravitational effects, which get amplified in
certain regimes ($r\ll\lambda$) as a result of the gravitational potential
correction (\ref{power potential}). We begin our analysis by considering the
perturbation as being described in terms of the gravitational Hamiltonian,%

\begin{equation}
H_{g}=m_{i}V\left(  r\right)  , \label{hamiltonian}%
\end{equation}
where $m_{i}$ is the lepton mass.

In order to obtain the constraints provided by this model
(\ref{power potential}), we will find the contribution of gravitational energy
to the energy levels by applying perturbation formalism. In this case, it is
important to stress that the eq. (\ref{hamiltonian}) corresponds to a small
contribution to the total atomic Hamiltonian. Under this assumption, one can
analytically determine the general expression for the gravitational energy
contribution on energy levels of hydrogen-like ion with $Z$ atomic number and
then express it as%

\begin{eqnarray}
\mathcal{E}_{g,i}\left(  n,l\right)\equiv\left\langle H_{g}\right\rangle
_{n,l} & = & - \frac{G_{N}m_{i}M}{a_{0,i}}\frac{Z}{n^{2}}\Bigg(1+_{3}%
F_{2}(n-l,1-\delta,-2l-\delta;-2l,n-l-\delta+1;1)\nonumber \\
&&\times\frac{(-2)^{\delta}\Gamma(2l+1)\Gamma(n-l)\,}{n\Gamma
(\delta)\Gamma(2l+\delta+1)\Gamma(n-l-\delta+1)}\left(  \frac{2Z\lambda
}{a_{0,i}n}\right)  ^{\delta}\Bigg)  , \label{general}%
\end{eqnarray}
where the average $\left\langle {}\right\rangle _{n,l}$ is calculated with
respect to well-known unperturbed wave functions for the hydrogen atom, $M\cong
Am_{p}$ is the nuclear mass, $a_{0,i}$ is the Bohr radius,\ $_{p}F_{q}\left(
a;b;c\right)  $ is the generalized hypergeometric function, and the $i$-index
labels the lepton orbiting the hydrogen-like ion, i.e., $i=e\left(
\mu\right)  $ for an electronic (muonic) ion.

Let us now consider a hydrogen-like ion with $Z$ atomic number found an
$n$-state with $l=n-1$ angular momentum. In this case, the gravitational
non-Newtonian potential energy, $E_{g,i}\left(  n\right)  =\mathcal{E}_{g,i}\left(  n,n-1\right)$, can be reduced to:%

\begin{equation}
E_{g,i}\left(  n\right) =-\frac
{G_{N}m_{i}M}{a_{0,i}}\frac{Z}{n^{2}}\left(  1+\frac{\Gamma\left(
2n-\delta\right)  }{\Gamma\left(  2n\right)  }\left(  \frac{2Z\lambda}%
{a_{0,i}n}\right)  ^{\delta}\right)  , \label{energy pert1}%
\end{equation}
where $\Gamma\left(  n\right)  =\left(  n-1\right)  !$ is the Gamma function.
The analytic expression (\ref{energy pert1}) provides us with the shift on energy
levels of the Rydberg atom in the $n$-state ($l=n-1$) due to the gravitational
potential energy contribution. We should note that the expected classical
result to $4$-dimensional gravitational energy due to the atomic nucleus is
recovered when $\lambda=0$. Through the eq. (\ref{energy pert1}), one can see
that gravitational energy is amplified at scale $\lambda>a_{0,i}n/2Z$, if the
gravitational potential has power-type correction.

\subsection{Constraints on power-law-type deviation%
\label{ssec:num1}%
}

A promising spectroscopic measurement technique involving optical frequency
combs proposes a method to measure frequency transitions with remarkably high
empirical precision. In this sense, the transition measurements between
Rydberg's states, localizing in the optical frequency range ($100$\textrm{THz}
- $1000$\textrm{THz}), will be performed with a promised relative uncertainty
of the order of $10^{-19}$\ \cite{opticalcomb,opticalmetrology}. Thus, the
limitation in the extraction of the Rydberg constant, due to high
uncertainties originated by the atomic structure, and which is evidenced in
the proton radius problem \cite{nature,science}, is circumvented
\cite{Jentschura}. In this way, a previous paper \cite{lemos3}, showed that
measurements of transitions between Rydberg states located in the optical
frequency spectrum of muonic ions would present measurable gravitational
effects in braneworld scenarios and, therefore, has demonstrated to be an
alternative way to search for extra dimensions empirical traces.

Following this idea, let us find the prospective constraints by considering
that transitions occur between adjacent states (($n,l=n-1$) and ($n-1,l=n-2$%
)), and, in the case of Rydberg states, will be limited by the general
expression (\ref{constraint1}). Thus, the gravitational contribution to energy
levels should be less than the promised empirical uncertainty for the leading
term contribution of the atomic Hamiltonian%

\begin{equation}
\Delta E_{g,i}\equiv\left\vert E_{g,i}\left(  n\right)  -E_{g,i}\left(
n-1\right)  \right\vert \leq\delta E_{\text{\textquotedblleft}\exp
\text{\textquotedblright}}, \label{constraint1}%
\end{equation}
where $\delta E_{\text{\textquotedblleft}\exp\text{\textquotedblright}}$ is
the empirical uncertainty promised for transition and can be written as%

\begin{equation}
\delta E_{\text{\textquotedblleft}\exp\text{\textquotedblright}}%
=10^{-19}hcR_{\infty}Z^{2}\left(  \frac{1}{n^{2}}-\frac{1}{\left(  n-1\right)
^{2}}\right)  ,
\end{equation}
and $R_{\infty}$ is the Rydberg constant, $h$ is the Planck constant, and $c$
is the speed of light. The eq. (\ref{constraint1}) will provide us the
theoretical constraints on the $\lambda$-parameter for certain $\delta$-values fixed.

Here, one can present the constraints on the gravitational contribution
(\ref{constraint1}) by analyzing the transition $n=15$ to $n=14$, for
instance, in the electronic and muonic neon ion ($^{20}\mathrm{Ne}^{+9}$) -
see, e.g., \cite{erickson1976} -, assuming that such contribution is of the
order of, or lower than the promised\ uncertainty of the leading term of this
transition involving Rydberg states. These selected states ensure that
frequency transitions lie about the optical band
\cite{opticalcomb,opticalmetrology}. For the sake of simplicity, we must still
mention that the muonic hydrogen-like ions are atoms whose lepton orbiting the
nucleus is the muon, e.g., see \cite{nature,science}.%

\begin{figure}
[h]
\begin{center}
\includegraphics[
height=2.3566in,
width=3.1964in
]%
{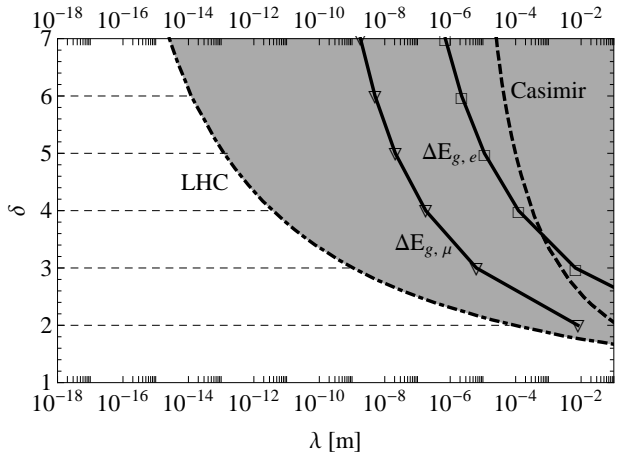}%
\caption{Constraints on parameters of the power-law gravitational potential.
The symbols \textquotedblleft$\triangledown$\textquotedblright\ $\left(
\text{\text{\textquotedblleft}}\square\text{\text{\textquotedblright}}\right)
$ indicate the fixed value\ to $\delta=2,3,4,5,6$ and $7$ in
muonic\ (electronic)\ Neon ion. These curves\ (LHC and Casimir), were obtained
from Fig. 11 of Ref. \cite{murata} using plot-digitizer software
\cite{digitizer}.}%
\label{fig1}%
\end{center}
\end{figure}

In fig. \ref{fig1}, one can observe that for $\delta>3$, the spectroscopic
constraints of the electronic ion ($\Delta E_{g,e}$ curve) are more
restrictive than the bounds obtained through the Casimir effect. In its turn,
the constraints found for the muonic ion transition spectroscopic data lead to
stronger bounds than those obtained from the Casimir physics, for all values of
$\delta$ considered.

Before proceeding, it is important to stress that although the constraints
found by spectroscopic analysis for this particular non-Newtonian potential
are weaker than those found by the LHC, this study presents another independent way of testing deviations to Newton's gravity law. For
more details on obtaining the LHC and Casimir curves, see Ref. \cite{murata}.

\section{Bounds from Yukawa-type potential}

Let us analyze another parametrization that describes deviations to
Newton's gravity law which has been motivated, at first, by theories
predicting a new long-range interaction of Yukawa-type. These deviations from
the predictions to gravitational potential have also been proposed in some
scenarios as extra-dimensional theories \cite{add1,add2,kehagias}, and even in
Casimir Physics \cite{mostepanenko,mostepanenko2,murata}, for instance.{} The
Yukawa parametrization presents a generalized form to the gravitational
potential generated by a mass M which can be written as:%

\begin{equation}
V_{Yu}\left(  r\right)  =-G_{N}\frac{M}{r}\left(  1+\alpha e^{-r/\lambda
}\right)  \equiv V_{N}\left(  r\right)  +\delta V_{Yu}\left(  r\right)  ,
\label{add potential}%
\end{equation}
where $G_{N}$ is the $4$-dimensional Newtonian gravitational constant,
$V_{N}\left(  r\right)  $ is the usual Newtonian gravitational potential, and
$\delta V_{Yu}\left(  r\right)  $ is the Yukawa correction\ to usual
potential. According to extra-dimensional scenarios, the $\lambda$-parameter
will describe the compactification scale of the higher-dimensional space,
which is assumed to be on the sub-mm scale to keep consistency with the
currently empirical data obtained. At the same time, $\alpha$ is a coupling
constant related to hidden spacetime dimensions \cite{kehagias}.

Such as we proceeded in the previous subsection
\ref{ssec:num1}%
, one will apply the perturbation method to calculate the gravitational
contribution to transition frequencies by considering atoms in Rydberg's
states. Therefore, one can consider the potential (\ref{add potential}) as a
perturbation to atomic energy levels and rewrite it in the Hamiltonian form
$\widetilde{H}_{g}=m_{i}V_{Yu}\left(  r\right)  $. In this case, we are
interested in calculating the non-Newtonian gravitational contribution for a
given energy level $n$ of a hydrogen-like atom in Rydberg's state ($n$ and
$l=n-1$), which can be evaluated analytically and thus yields the expression, $\widetilde{E}_{g,i}\left(  n\right)  \equiv\left\langle \widetilde{H}%
_{g}\right\rangle _{n,l}$,%

\begin{equation}
\widetilde{E}_{g,i}\left(  n\right)=-\frac{G_{N}m_{i}M}{a_{0,i}}\frac{Z}{n^{2}}\left[
1+\alpha\left(\frac{2\lambda Z/na_{0,i}}{
1+2\lambda Z/na_{0,i}}\right)  ^{2n}\right]  . \label{energy pert2}%
\end{equation}

We emphasize that for the $\alpha\rightarrow0$, one can recover the expected
four-dimensional behavior for the gravitational potential energy. In this
case, the $\alpha$ parameter turns off the non-Newtonian effects generated by
the Yukawa correction.

\begin{figure}
[h]
\begin{center}
\includegraphics[
height=2.3557in,
width=3.5812in
]%
{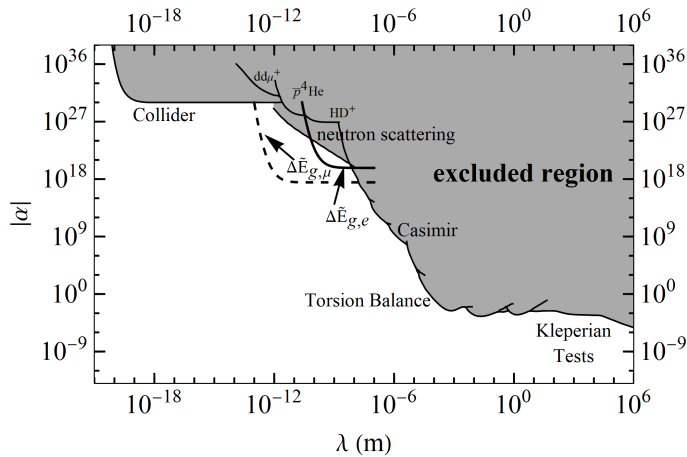}%
\caption{Prospective bounds obtained by the non-Newtonian Yukawa-type
potential parametrization. The curve $\Delta\widetilde{E}_{g,e}$ describes the
limits obtained considering the electronic neon ion transition, while $\Delta\widetilde
{E}_{g,\mu}$ is found from the muonic ion transition (neon).}%
\label{fig2}%
\end{center}
\end{figure}

Finally, we will find the bounds on free parameters $\alpha$ and $\lambda$
from the study of transitions involving adjacent states by an analogous
procedure to that of the previous subsection
\ref{ssec:num1}%
. Thus, admitting that the gravitational contribution to energy levels on
transitions in hydrogen-like Rydberg atoms must be less than promised
empirical uncertainty, one obtains the constraints on parameters from Yukawa
corrected potential. This condition ensures that no traces of new physics have
been detected so far. For this study, we analyze the transition from $n=14$ to
$n=15$ for the electronic and muonic ion of Neon ($^{20}\mathrm{Ne}^{+9}$),
and we find the bounds for the parameters $\lambda$ and $\alpha$, see fig.
\ref{fig2}.%

The gravitational contribution to transition energy has been provided by the equation $\Delta\widetilde{E}_{g,i}\equiv\left\vert \widetilde{E}_{g,i}\left(
n\right)  -\widetilde{E}_{g,i}\left(  n-1\right)  \right\vert $. It is
important to highlight that the non-Newtonian gravitational energy will
increase the energy gap between adjacent states with the principal quantum
numbers $n$ and $n-1$. In fig. \ref{fig2} we have compared the obtained bounds
from this study with a wide variety of empirical constraints originated from
different sources such as neutron scattering, spectroscopy of exotic atoms,
collider, see, e.g., \cite{murata,salumbides,neutron,Antoniadis}. From Ref.
\cite{murata}, the collider data have been extracted using plot-digitizer
software \cite{digitizer}. The additional empirical bounds can be found in
Ref. \cite{salumbides}. As is seen, the constraints obtained for
$\lambda<10^{-8}\mathrm{m}$ are stronger for both cases of Rydberg state
transitions (electronic and muonic) than the bounds obtained through the
Casimir effect analysis. The shaded areas represent the excluded region for
the existence of non-Newtonian effects.

At last, let us emphasize that, although we have obtained
stronger bounds as they are prospective constraints, one may hope that these
results will be weakened since the empirical data, together with their
respective uncertainty ($\delta E_{\exp}$), are provided. By taking into
account, the theoretical uncertainty ($\delta E_{th}$) obtained by study all
terms of atomic Hamiltonian, one also expects a weakening of the obtained
constraints. In this case, the new constraints can be found by using the
combined uncertainty, defined by $\delta E=\sqrt{\delta E_{th}^{2}+\delta
E_{\exp}^{2}}$, to limit non-Newtonian deviations on atomic energy levels.

\section{ Non-Newtonian parametrizations constraints comparison}

As discussed in the preceding sections, the inverse square law corrections are
proposed in many scenarios considering different parametrizations. In
principle, the different proposed deviations to Newton's gravitational law
would produce unlike effects on the atomic energy levels. In turn, it is
possible, for instance, to compare the gravitational energy contribution with
the leading term contribution to the energy levels of Rydberg atoms
$E_{n}=-cR_{\infty}Z^{2}/n^{2}$. In this case, the expected deviation for
atomic energy levels due to the non-Newtonian contribution is estimated, as
discussed, e.g., in \cite{ryd}, where a new Yukawa-type interaction has been considered.

\begin{figure}
[h]
\begin{center}
\includegraphics[
height=2.2857in,
width=3.6037in
]%
{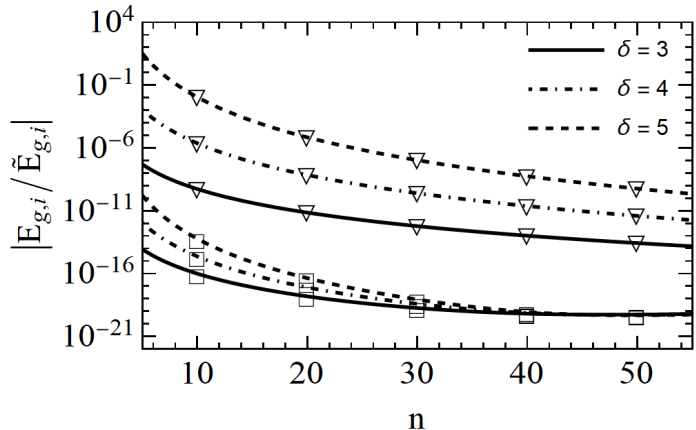}%
\caption{Estimation of the ratio between non-Newtonian effects due to
power-law-type ($E_{g,i}$) and Yukawa-type ($\tilde{E}_{g,i}$)
parameterizations on the atomic energy levels. The \textquotedblleft$\square
$\textquotedblright\ symbol identifies the electronic atom while the
\textquotedblleft$\triangledown$\textquotedblright\ symbol labels the muonic
atom.}%
\label{fig3}%
\end{center}
\end{figure}

In another way, we aim to compare the effects produced by Yukawa-type
potential and power-law potential on the energy levels of muonic and
electronic atoms. In this case, we study the relationship between the effects
produced in the two proposed scenarios, $E_{g,i}\left(  n\right)  /\tilde
{E}_{g,i}\left(  n\right)  $, considering hydrogen-like ions in Rydberg states
($l=n-1$). As shown in fig. \ref{fig3}, the deviations found for the two
studied parametrizations have contributed distinguishably to atomic energy
levels. Therefore, if such deviations are measured, we could, in principle,
determine the parametrization that best fits the empirical data and,
hence, predicts the possible origin of the inferred correction for the
atomic levels.%

To obtain fig. \ref{fig3} we have considered an hydrogen-like ion with $Z=10,$
$\lambda=10^{-8}\mathrm{m},$ and $\alpha=10^{20}$. The corrections originated
from the Yukawa parametrization showed to be stronger than those obtained by
power-law-type correction, except in the case of muonic atoms in states with
$n<10$ e $\delta>4$. We also infer that for high values {}{}of $n$ $\left(
n\gg50\right)  $, the correction produced is approximately $\delta
$-independent, as seen in the electronic atom. We notice that the effects on
muonic atoms are amplified compared to electronic atoms, mainly due to the
smaller Bohr radius of the muonic system ($a_{0,\mu}$). Thus, spectroscopy of
muonic atoms has been shown as an exciting approach in searching for new physics.

\section{Concluding Remarks}

Summarily we have discussed the deviations of Newton's inverse-square law
proposed from several scenarios. The fifth force theory, for instance,
predicts Newton's gravitational law deviations revealed by a new long-range
interaction on the submillimeters scale. Furthermore, in Casimir physics, the claimed deviations to Newtonian gravity have been studied and would be manifest in short-distance
scaling experiments due to the exchange of light or massless particles via an
interaction of the Yukawa-type or power-law-type. On the other hand, recent
higher-dimensional theories postulate in their description the existence of
modifications of the usual $(3+1)$-dimensional Newtonian potential. In known
braneworld models, deviations from inverse-square law originate from the
existence of hidden spatial dimensions. According to such models, the
Newtonian gravitational potential would take the form of Yukawa potential at a
short-range scale.

In this work, we have explored the possibility of making a theoretical
prediction on deviations of Newtonian gravity. Thus, one have presented
constraints by considering the Yukawa and power-law parametrizations as
corrections to Newton's gravitational law. We have calculated deviations to
energy levels of electronic and muonic ions in Rydberg states due to
gravitational potential energy contribution corrected by the Yukawa and
power-law-type parametrizations. Then, we presented and discussed the bounds
obtained for the free parameters of both models ($\delta,\alpha,$ and
$\lambda$), comparing them with constraints from different origins.

Although the bounds found in this analysis are stronger than results obtained
from the Casimir force measurements, it is important to stress that we found prospective constraints in this study. For the bounds obtained through
the study of power-law potential, we find stronger constraints for
muonic ion than the results of the electronic ion and data from the Casimir
physics. As shown, the Yukawa-type potential presents stronger bounds to
$\lambda<10^{-8}\mathrm{m}$, excluding any deviations to Newton's
inverse-square law to $\alpha>10^{20}$. For this analysis, we required that
the gravitational energy contribution corrected by the Yukawa-type and
power-type parametrizations does not exceed the theoretical uncertainty of the
leading term of atomic Hamiltonian, ensuring relative promised uncertainty to be of order $10^{-19}$. In turn, the uncertainty has been predicted by spectroscopy
techniques of high-precision optical frequency combs.

Finally, we compare the deviations on Rydberg ion energy levels generated from
the two parametrizations studied. In this case, considering a hydrogen-like
ion with $Z=10$, we found that the Yukawa-type gravitational energy
contribution is stronger than the correction obtained for the power-law-type
potential for $n>10$ and $\delta\leq4$ (fixing $\lambda=10^{-8}\mathrm{m}$,
and $\alpha=10^{20}$). The detecting deviations in ionic energy levels would
indicate a fifth force of nature. Since the discussed corrections are
described for different regimes and proposed in unlike scenarios, from the
measurement deviations, it would be possible to infer, in principle, the
origin of deviations.

\begin{acknowledgments}
I would like to thank the referees for their suggestions and valuable comments. ASL is supported by Conselho Nacional de Desenvolvimento Científico e Tecnológico - Brasil (CNPq) through the project No. 150601/2021-2.
\end{acknowledgments}

\end{document}